\documentclass[letterpaper,12pt]{article}
\newtheorem{theorem}{Theorem}[section]

\newtheorem{corollary}[theorem]{Corollary}

\newenvironment{proof}[1][Proof]{\begin{trivlist}
\item[\hskip \labelsep {\bfseries #1}]}{\end{trivlist}}
\newenvironment{definition}[1][Definition]{\begin{trivlist}
\item[\hskip \labelsep {\bfseries #1}]}{\end{trivlist}}

\newenvironment{remark}[1][Remark]{\begin{trivlist}
\item[\hskip \labelsep {\bfseries #1}]}{\end{trivlist}}

\newcommand{\qed}{\nobreak \ifvmode \relax \else
      \ifdim\lastskip<1.5em \hskip-\lastskip
      \hskip1.5em plus0em minus0.5em \fi \nobreak
      \vrule height0.75em width0.5em depth0.25em\fi}

\usepackage{amsmath,amssymb}
\numberwithin{equation}{section}
\title{\bf\large
Convex Risk Measures: Lebesgue Property on one Period and Multi Period Risk Measures and Application in Capital Allocation Problem}
\author{Hirbod Assa\thanks{Pavillon Andr\'e-Aisenstadt 2920, chemin de la Tour, bureau 5190
Montreal, Quebec H3T 1J4}\\ 
\tt assa@dms.umontreal.ca \\
\small Department of Mathematics and Statistics \&\\
\small Group for Research in Decision Analysis (GERAD)\\
\small University of Montreal\\}
\date{}

\begin{document}

\maketitle

\abstract{In this work we study the Lebesgue property for convex risk measures on the space of bounded c\`adl\`ag  random processes ($\mathcal{R}^\infty$). Lebesgue property has been defined for one period convex risk measures in \cite{Jo} and earlier had been studied in \cite{De} for coherent risk measures. We introduce and study the Lebesgue property for convex risk measures in the multi period framework. We give presentation of all convex risk measures with Lebesgue property on bounded  c\`adl\`ag processes. To do that we need to have a complete description of compact sets of $\mathcal{A}^1$. The main mathematical contribution of this paper is the characterization of the compact sets of $\mathcal{A}^p$ (including $\mathcal{A}^1$). At the final part of this paper, we will solve the Capital Allocation Problem when we work with coherent risk measures.
 }

\section{Introduction}
Assessing and qualifying risk is important in many aspects of human activities. It seems rational to expect something "good" by accepting something "bad" but the question is how much we should be rewarded. Is it "good enough" to accept the risk? The other question is among several risk exposures how to choose the one that represent a smaller risk .

Markowits ,economist, worked to make explicit the trade-off of the risk and reward in the context of portfolio of financial assets,see \cite{Ma}, \cite{Sh} and \cite{Li}. He has already considered that the returns distribution of assets is jointly Normal (or Gaussian).

The variance is a  risk measure which assesses both sides of the Profit and Loss distribution. That means variance takes into account the losses as well as profits. Regards this problem and other technical problems the new measure, Value at Risk, was defined and popularized in 1994 :
$$VaR_\alpha(X)=-\inf\{x\vert\mathbb{P}[X\leq x]>\alpha\}$$
Value at Risk (VaR) measures the left tail of P\& L random variable which is connected with the Losses. In other words VaR is a measure showing how the market value of an asset or of a portfolio of assets is likely to decrease over a certain time period, under usual conditions. It is typically used by security houses or investment banks to measure the market risk of their asset portfolios (market value at risk), but is actually a very general concept that has broad application. As references one can consult \cite{Cr},\cite{Gl},\cite{Ho},\cite{Jo}. 

Unlike variance, VaR does not meet the sub additivity property. Regarding to this problem, in 1998 Artzner et al\cite{Ar} proposed an axiomatic way to define a generation of risk measures called coherent risk measures. Their idea is to define a risk which can be used as capital requirements, to regulate the
risk assumed by market participants, traders, insurance underwriters, as well as to
allocate existing capital. They propose the following definition \cite{Ar}: 
\begin{definition}. A function $\rho:L^\infty\rightarrow \mathbb{R}$ is
 a Coherent Risk measure if 
\item[1-]$\rho(\lambda X+(1-\lambda)Y)\leq \lambda\rho(X)+(1-\lambda)\rho(Y)$ for any $X,Y\in
 L^\infty$ and $\lambda\in[0,1]$.(Convexity)
\item[2-]$\rho(\lambda X)=\lambda \rho(X)$ for any $X\in L^\infty$ and
 $\lambda>0$.(Positive Homogeneity)
\item[3-]$\rho(X+m)=\rho(X)-m$ for any $X\in L^\infty$ and
 $m\in\mathbb{R}$.(Translation Invariant)
\item[4-]$\rho(X)\leq\rho(Y)$ $\forall X,Y \in L^\infty$ and $Y\leq
 X$.(Decreasing) 
\end{definition}
Condition 1 can be replaced by subadditivity condition, $\rho(X+Y)\leq\rho(X)+\rho(Y)$.
Each condition stated in responds to economic reasons but they remain somehow controversial.
In their seminal work \cite{Ar} the authors also define the new risk measure, Expected Shortfall, as a particular example of coherent risk measures. Expected Shortfall measures the expectation of losses below Value at Risk. The definition is as follows:
$$ES[X]=E[X\vert X<VaR_\alpha(X)].$$ 

Later F\"ollmer et al \cite{Fo} proposed to generalize the coherent risk in a natural way to convex risk measure by relaxing the condition 2. See for example \cite{Fo}. 

The main subject of \cite{De} and \cite{Fo} is the characterization of risk measures with the so-called Fatou property (see next section). Actually Fatou property is counterpart of the concept of lower semi continuity in weak star topology. The other property which seems essential to study the risk measures is the so-called Lebesgue property (see next section). This property first time has studied under this name for convex risk measures in \cite{Jo}, but earlier this property had been studied for coherent risk measures in \cite{De}. The author of \cite{De} showed how the coherent risk measures with Lebesgue property can solve the Capital Allocation Problem (see section 4). To study the Lebesgue property we need to know the complete description of weak compact sets of $L^1$. 

After some discussion on risk measures for random variables, naturally the next step is to define the risk measures for random processes. Artzner et al in \cite{Ar2} gave a simple way to generalize the coherent risk measures for discrete time models. For the first time the convex risk theory was defined and studied for continuous time random process in \cite{Ch}. The authors of \cite{Ch} defined the coherent and convex Risk measure theory for bounded c\`adl\`ag processes. Then they extend the theory of convex risk measures on the space of  unbounded c\`adl\`ag  processes in \cite{Ch2}. The main subject of \cite{Ch} is restating and characterizing the Fatou property forrisk measures on  bounded c\`adl\`ag  processes. Unlike the Fatou property, Lebesgue property has not define and studied yet and our main subject in this paper is to define and study Lebesgue property for bounded c\`adl\`ag  processes. Actually the main mathematical contribution of this paper is the characterization of the weak compact sets of $\mathcal{A}^1$ (see section 2.2) which is assumed as the dual space of c\`adl\`ag  processes. The main application of our result is finding the solution of Capital Allocation Problem in the c\`adl\`ag  processes framework.

The paper is organized as follows: in section 2, within two subsections we give the preliminarily definitions and results for one period and multi period risk measures. In section 3 we give some definitions and remarks which are needed for our results in section 4. In section 4 we give our main mathematical contributions of this paper. In section 5 we discuss the Capital Allocation Problem in the framework of c\`adl\`ag  processes. Section 6 is devoted to conclusions and finally in appendix we give the proof of theorems 4.2.

\section{One period and multi period Convex Risk Measures }

In this section we briefly review the concepts which preliminarily should be studied. In the first subsection we give preliminary definitions and results on one period convex risk measure. In the second subsection our subject is to give the same definitions of subsection 1 for multi period convex risk measures and we study the same results on the space of random processes. 
\subsection{One period risk}
We give the following definition from \cite{De} and \cite{Jo}: 
\begin{definition}:A convex risk $\rho$ is has Fatou property if for any bounded sequence $X_n$ in $L^\infty$ which converges in probability to $X$ then:
$$\rho(X)\leq\liminf\rho(X_n).$$
and we say $\rho$ has Lebesgue property if always the equality occurs. 
\end{definition}
Fatou property seems to find its source in the classical theory of locally solid Riesz spaces with Fatou topologies\cite{Al}.
Actually the following theorem shows that Fatou property is defined as the probabilistic counterpart of the concept of lower semi continuity in weak star topology \cite{Fo}. 

\begin{theorem}:Let $\rho:L^\infty\longrightarrow \mathbb{R} $ be a convex risk measure and let $P$ be the set of measures $\mathbb{Q}$ such that $\mathbb{Q}\ll \mathbb{P}$. Then the following are equivalent:
\item[(i)] The function $\rho$ has Fatou property .
\item[(ii)]There is a penalty function $\alpha:P\rightarrow (-\infty,+\infty]$ such that:
\begin{equation}
\label{represent}
\rho(X)=\sup\limits_{\mathbb{Q}\in P}\{-E^\mathbb{Q}[X]-\alpha(\mathbb{Q})\}.
\end{equation}
\end{theorem}
From the theory of convex functions we know that one can get $\alpha=\rho^*$ where $\rho^*$ is the conjugate function:
\begin{equation}
\label{conjugate}
\rho^*(\mathbb{Q})=\sup\limits_X\{E^\mathbb{Q}[X]-\rho(X)\}.
\end{equation}
By the theory of convex functions and relation \ref{represent} we know that for any $c\in\mathbb{R}$, the contour set $\{\rho\leq c\}$ is convex and weak* closed.

On the other hand the contour set $\{\rho^*\leq c\}$ is always weakly closed subset of $L^1$. The Lebesgue property gives more information about this contour set. Actually we have the following theorem from \cite{Jo}:
\begin{theorem}Let $\rho:L^\infty(\Omega)\longrightarrow \mathbb{R}$ be a risk measure. The following conditions are equivalent:
\\
\item[1-]$\rho$ has Lebesgue property
\\
\item[2-]$\{\mu\in L^1_+ | \rho^*(\mu)\leq c\}$ is $\sigma(L^1,L^\infty)$-compact subset of $L^1$  for every $c\in\mathbb{R}^+$.
\label{th:Jo}
\end{theorem}  

If we want to restate this theorem for coherent risk measures we should say $\rho$ has Lebesgue property iff the set $\{\mu\vert\rho^*(\mu)=0\}$ is weakly compact.
\begin{remark}.
From Dunford-Pettis Theorem we know that a subset of $L^1$ is relatively weak compact iff it is uniformly integrable. That means we could say $\{\mu\in L^1_+ | \rho^*(\mu)\leq c\}$ is uniformly integrable, instead of weak compact. 
\end{remark}

\subsection{Multi period risk}

In the real world the problems are in the time periods and the fluctuation of the financial or economical variables during these periods are important.
To have a measure that can measure the risk of "random processes" we should restrict the study on a special spaces of random processes. On the other hand when we work with the random processes the flow of information (i.e. filtration) should be taken into account.
In the sequel we assume that all objects are defined in a standard probability space without any atom $(\Omega,\mathcal{F},\mathbb{P})$. This space is endowed with a filtration $(\mathcal{F}_t)_{0\leq t\leq T}$ satisfying the usual conditions. Let $\mathcal{R}^q$ be as follows:
\begin{equation} 
 \mathcal{R}^q=\left \{X:[0,T]\times\Omega\longrightarrow \mathbb{R}\Bigg |
\begin{array}{clcr}
& X \,\,\text{is c\`adl\`ag} \\
& X \,\,\text{adapted}\\
& (X)^*\in L^q
\end{array}
 \right\} ,
\end{equation} 
This space is equipped with the following norm:
\begin{equation}
\Vert X\Vert_{\mathcal{R}^q}=\Vert X^*\Vert_{q},
\end{equation}
where $X^*(\omega)=\sup\limits_{0\leq t\leq T}\vert X_t(\omega)\vert\,\,,\,\,\omega\in\Omega$.

In \cite{Ch}, the authors have suggested to use the space of c\`adl\`ag  bounded processes as the space of risky financial items. They also suggest to use the space $\mathcal{A}^p$ as dual. This space is defined as follows:
 \begin{equation} 
\left \{a:[0,T]\times\Omega\longrightarrow \mathbb{R}^2\Bigg |
\begin{array}{clcr}
& a=(a^{pr},a^{op}),\\ 
& a^{pr},a^{op} right\,\text{continuous ,finite\,variation}\\
& a^{pr} \text{predictable},a^{pr}_0=0\\
& a^{op}\text{optional , purely\,discontinuous}\\
& \text{Var}(a^{pr})+\text{Var}(a^{op})\in L^p
\end{array}
 \right\} \;,
\end{equation}

where $\text{Var}(f)$ is the variation of function $f$. The space $\mathcal{A}^p$ is equipped with the following norm:
\begin{equation}
\Vert (a^{pr},a^{op})\Vert_{\mathcal{A}^p}=\Vert \text{Var}(a^{pr})+\text{Var}(a^{op})\Vert_{L^p}.
\end{equation}
The dual relation between $\mathcal{A}^p,\mathcal{R}^q$ is defined as:
\begin{equation}
\langle X,a\rangle=E[(X|a)],
\end{equation}
where $(X|a)=\int\limits_0^TX_{t-}da_t^{\text{pr}}+\int\limits_0^TX_tda_t^{\text{op}}.$
\\Again following \cite{Ch}, let
\begin{equation}
\mathcal{D}_{\sigma}:=\{a\in\mathcal{A}^1_+;\Vert a\Vert=1\},
\end{equation}
where $\mathcal{A}^1_+=\{a=(a^{pr},a^{op})\in\mathcal{A}^1 \vert a^{pr},a^{op}\,\,\, \text{ are non-decreasing}\}$
\\

Before moving on with our discussion we need to lay down a few definitions.

\begin{definition}
A  convex risk measure is a function $\rho:\mathcal{R}^\infty\longrightarrow  \mathbb{R}$ in which : 
\\
\item[1-]$\rho(\lambda X+(1-\lambda)Y)\leq \lambda\rho(X)+(1-\lambda)\rho(Y)$ for any $X,Y\in
 \mathcal{R}^\infty$ and $0\leq \lambda\leq1$.(Convex)
\item[2-]$\rho(X+m)=\rho(X)-m$ for any $X\in \mathcal{R}^\infty$ and
 $m\in\mathbb{R}$.(Translation Invariant)
\item[3-]$\rho(X)\leq\rho(Y)$ $\forall X,Y \in \mathcal{R}^\infty$ and $Y\leq
 X$.(Decreasing)
\\
\\
We call it coherent if in addition:
\item[4-]$\rho(\lambda X)=\lambda \rho(X)$ for any $X\in \mathcal{R}^\infty$ and
 $\lambda>0$.(Positive Homogeneous) 
\end{definition}

In \cite{Ch}, the authors give the following definition of Fatou property for risk measures on $\mathcal{R}^\infty$. 
\begin{definition}
The convex function $\rho$ has Fatou property if for any bounded sequence $\{X_n\}_{n\in\mathbb{N}}\subseteq \mathcal{R}^\infty$, in which for some $X\in \mathcal{R}^\infty$, $(X_n-X)^*\xrightarrow{\mathbb P} 0$, we have $\rho(X)\leq\liminf\rho(X_n)$.  
\end{definition}

From \cite{Ch}, we have the following characterization :

\begin{theorem}
Let $\rho:\mathcal{R}^\infty\longrightarrow \mathbb{R}$ be a risk measure. The following are equivalent:
\\
\item[1-] $\rho$ is represented as:
\begin{equation}\label{2.6}
\rho(X)=\sup\limits_{a\in\mathcal{D}_\sigma}\{-\langle X,a\rangle-\gamma(a)\}\,\,,\,\,X\in \mathcal{R}^\infty
\end{equation} 
where $\gamma$ is a so-called penalty function $\gamma:\mathcal{D}_\sigma\rightarrow (-\infty,+\infty]$ such that $-\infty<\inf\limits_{a\in\mathcal{D}_\sigma}\gamma(a)<\infty$.
\\
\item[2-]$\rho$ is a convex risk measure on $\mathcal{R}^\infty$ such that $\{X\in\mathcal{R}^\infty|\rho(X)\leq0\}$ is $\sigma(\mathcal{R}^\infty,\mathcal{A}^1)$-closed.
\\
\item[3-]$\rho$ has Fatou property.
\\
\item[4-]$\rho$ is continuous for bounded increasing sequences.
\\
\\Moreover, in each case, the conjugate function $\rho^*$, restricted to $\mathcal{D}_\sigma$, is a penalty function which is bigger than $\gamma$ and $\gamma$ can be replaced by $\rho^*$ in the first expression above.
\end{theorem}

Later on, in \cite{Jo} we find the notion of Lebesgue property for risk measures on $L^\infty(\Omega)$. In this paper we aim to characterize this property for risk measures on $\mathcal{R}^\infty$ and so we need the following extended definition:

\begin{definition}
The convex function $\rho$ has Lebesgue property if for any bounded sequence $\{X_n\}_{n\in\mathbb{N}}\subseteq \mathcal{R}^\infty$, in which for some $X\in \mathcal{R}^\infty$, $(X_n-X)^*\xrightarrow{\mathbb P} 0$, we have $\rho(X)=\lim\rho(X_n)$.
\end{definition}

In \cite{Jo}, we find the following characterization theorem for convex risk measures on $L^\infty(\Omega)$:


\begin{theorem}Let $\rho:L^\infty(\Omega)\longrightarrow \mathbb{R}$ be a risk measure. The following conditions are equivalent:
\\
\item[1-]$\rho$ has Lebesgue property
\\
\item[2-]$\{\mu\in L^1_+ | \rho^*(\mu)\leq c\}$ is $\sigma(L^1,L^\infty)$-compact subset of $L^1$  for every $c\in\mathbb{R}^+$.
\label{th:Jo}
\end{theorem}

One of the main contributions of this paper is to give a characterization theorem analogous to Theorem \ref{th:Jo} for risk measures on $\mathcal R^{\infty}$. 
 
\section{Static Risk and Some Remarks}
In this section we give some definitions that are needed in Section 4.

 Consider $\hat {\mathcal{F}_t}=\mathcal{F}$ and $(\hat {\mathcal{R}}^q,\hat {\mathcal{A}}^p)$ are the corresponding process spaces. Let $\Pi^{\text{op}}$ and $\Pi^{\text{pr}}$ be the optional and predictable projections. We also show the dual optional and predictable projection of finite variation processes with the same notation (as references see \cite{Ch},\cite{DM2} and \cite{Ka}). We define the projection $\Pi^*:\hat{\mathcal{A}}^p\rightarrow \mathcal{A}^p$ as follows : let $a=(a^l,a^r)\in\hat{\mathcal{A}}^p$ . Let $\tilde a^l=\Pi^{\text{pr}}(a^l)$ and $\tilde a^r=\Pi^{\text{op}}(a^r)$. Then one can split $\tilde a^r$ uniquely to purely discontinuous finite variation part $\tilde a^r_d$ and continuous finite variation part $\tilde a^r_c$ with $\tilde a^r_c(0)=0$. Now Define:
$$\Pi^*(a)=(\tilde a^l+\tilde a^r_c,\tilde a^r_d).$$
 We know that every predictable process is also optional so $\tilde a^l,\tilde a^r_c,\tilde a^r_d$ all are optional. This fact by  definition of $\Pi^*$ give that for every $X\in\mathcal{R}^q$ we have:
\begin{equation}
\label{3.0}
\langle X,a\rangle=\langle X,\Pi^*(a)\rangle.
\end{equation}
For more details see  relation 3.5 , Remark 3.6 \cite{Ch}.
\\Let $a=(a^{\text{pr}},a^{\text{op}})\in{\mathcal{A}}^p$. Since any predictable process is optional then by Theorem $2.1.53$ \cite{Ka} the measure $\mu(A)=\langle 1_A,a\rangle$ is optional and then $\langle X,a\rangle=\langle \Pi^{\text{op}}(X),a\rangle.$ This relation with \ref{3.0} give that $\forall X\in\hat{\mathcal{R}}^q,a\in\hat{\mathcal{A}^p}$:
\begin{equation}
\label{3.0.1}
\langle \Pi^{\text{op}}(X),a\rangle=\langle \Pi^{\text{op}}(X),\Pi^*(a)\rangle=\langle X,\Pi^*(a)\rangle.
\end{equation}  
For more details reader is referred to \cite{Ch} Remark 2.1 and Remark 3.6 and \cite{Ka}.

For every random variable $X\in L^q(\Omega,\mathcal{F})$ , we identify the constant random process $X_t:=X$ and $X$.
\\\textbf{Remark}{\bf 3.1}:Relation \ref{3.0} (or \ref{3.0.1}) shows that $\Pi^*$ is $\sigma(\hat{\mathcal{A}}^p,\hat{\mathcal{R}}^q)$/$\sigma({\mathcal{A}}^p,{\mathcal{R}}^q)$ continuous.
\\\textbf{Remark}{\bf 3.2}:Let $X\in L^q(\Omega)$ be a random variable. By Doob's Stopping Theorem
 it is easy to see that the optional projection of constant random process $X$ is the martingale $M_t:=E[X|\mathcal{F}_t]$.
 So then $\forall X\in L^q, a\in\hat{\mathcal{A}}^p_+$:
\begin{equation}
\label{3.2}
E[\text{Var}(a)X]=\langle X,a\rangle=\langle \Pi^{\text{op}}(X),a\rangle\,,\,\,
\end{equation}
\\Following  the paper \cite{Ch} define $\hat \rho=\rho\circ\Pi^{\text{op}}$. 
\begin{definition}
For every convex function $\rho$ on $\mathcal{R}^q$ the static convex function due to $\rho$ is defined on $L^q(\Omega,\mathcal{F})$ as follows:
$$\bar\rho(X):=\hat {\rho}(X)\,\,\,,\,\,\forall X\in L^q(\Omega,\mathcal{F}).$$
\end{definition}
\textbf{Remark}{\bf {3.3}}:As one can see the static convex functions due to $\rho$ and $\hat\rho$ are the same. On the other hand the arguments in the proof of Theorem 3.1 \cite{Ch} shows that if $(X_n-X)^*\xrightarrow{\mathbb{P}}0 $ then $(\Pi^{\text{op}}X_n-\Pi^{\text{op}}X)^*\xrightarrow{\mathbb{P}}0 $. So $\rho$ has Lebesgue property iff $\hat\rho$ has.\\ 
\textbf{Remark}{\bf {3.4}}By Theorem 2.1 every coherent risk measure could be identified with a subset $\mathcal{P}$ of $\mathcal{D}_\sigma$. Let $A=\text{Var}(\mathcal{P})$. Then by relation \ref{3.2} it is easy to see that:
\begin{equation}
\label{3.3}
\bar{\rho}(X)=E_A[-aX].
\end{equation} 
\section{Lebesgue Property of Risk Measures}
 By the abbreviation r.c. we mean relatively compact.
\begin{theorem}
Let $\rho:\mathcal{R}^\infty\longrightarrow \mathbb{R}$ be a convex function in which $\rho(-X)$ is a convex risk with Fatou property. Then the following are equivalent:
\\
\item[1-]$\rho$ has Lebesgue property.
\\
\item[2-]$\forall c\in\mathbb{R}^+$ the set $\{a\in\mathcal{A}^1_+\vert\rho^*(a)\leq c\}$ is  r.c. for topology $\sigma(\mathcal{A}^1,\mathcal{R}^\infty)$ .
\\
\item[3-]$\bar\rho$ has Lebesgue property.
\\
\item[4-]$\forall c\in\mathbb{R}^+$ the set $\{f\in L^1_+\vert\bar\rho^*(f)\leq c\}$ is  r.c. for topology $\sigma(L^1,L^\infty)$ .
\end{theorem}
Before giving the proof we give the following Theorem which is one of the most important contributions of thia paper. The proof is given in the Appendix.
\begin{theorem}
Let $A\subset\mathcal{A}^p$ and $\frac{1}{p}+\frac{1}{q}=1$. The following three conditions are equivalent:
\\1-$A$ is r.c. in the topology $\sigma(\mathcal{A}^p,\mathcal{R}^q)$. 
\\2-$\text{Var}(A)$ is r.c. in the topology $\sigma(L^p,L^q)$.
\\3-$C:=\{a_T-a_0|a\in A\}$ is r.c. in the topology $\sigma(L^p,L^q)$.
\end{theorem}

In addition we have two following Corollary.

\begin{corollary}
Let $A\subseteq\mathcal{A}^1$. Then $A$ is $\sigma(\mathcal{A}^1,\mathcal{R}^\infty)-$ r.c. iff $\text{Var}(A)$ is uniformly integrable.
\end{corollary}
\begin{corollary}
The set $A\subseteq\mathcal{A}^p$, for $p\neq\infty$, is $\sigma(\mathcal{A}^p,\mathcal{R}^q)-$ r.c. iff it is sequentially r.c. 
\end{corollary}

\textbf{Proof of theorem 4.1 }
\\(1)$\Rightarrow $(3). It comes out from definition. 
\\(3)$\Rightarrow $(4). Just Theorem 2.2.
\\(4)$\Rightarrow $(2). Let $a\in\mathcal{A}^1_+$ be such that $\rho^*(a)\leq c$ for some positive number $c$. Then by definition of conjugate function $\forall X\in\mathcal{R}^\infty$ we have $\langle X,a\rangle-\rho(X)\leq c$. Particularly this is true for every random process like $\Pi^{\text{op}}(X)$ where $X\in L^\infty$. By \ref{3.2} we get $E[\text{Var}(a)X]-\bar\rho(X)\leq c$. So we have $\text{Var}(\{a\in \mathcal{A}^1_+|\rho^*(a)\leq c\})\subseteq \{\mu\in L^1_+|\bar\rho^*(\mu)\leq c\}$. That means $\text{Var}(\{a\in \mathcal{A}^1_+|\rho^{*}(a)\leq c\})$ is  r.c. for topology $\sigma(L^1,L^\infty)$ and by Theorem 4.2 $\{a\in \mathcal{A}^1_+|\rho^{\,\,\,\,\,*}(a)\leq c\}$ is r.c. for topology $\sigma(\mathcal{A}^1,\mathcal{R}^\infty)$.
\\(2)$\Rightarrow $(1). First we consider that $\rho$ is positive homogeneous.
By this assumption, for every real number $c>0$, the set $\{a\in\mathcal{A}^1_+|\rho^*(a)\leq c\}$ is equal to $\{a\in\mathcal{A}^1_+|\rho^*(a)=0\}=:A$.
\\Let $X_n$ be a bounded sequence in $\mathcal{R}^\infty$ for which for some $X\in\mathcal{R}^\infty$ , $(X_n-X)^*\xrightarrow{\mathbb{P}}0 $. Since $\rho$ is positive homogeneous (then sub additive) and increasing we have :
$$\vert\rho(Z)-\rho(Y)\vert\leq\rho(Z-Y)^++\rho(Y-Z)^+\,\,,\,\,\forall Z,Y\in\mathbb{R}^\infty.$$
By this relation we could consider $X_n\geq0$ , $X=0$ and $(X_n)^*\xrightarrow{\mathbb{P}}0$. By the hypothesis (2) ,$A$ is r.c. for topology $\sigma(\mathcal{A}^1,\mathcal{R}^\infty)$. So by Lemma 4.2 the closed convex set $\text{Var}(A)$ is $\sigma(L^1,L^\infty)$-compact and as a consequence (by Theorem 2.2) the convex function $X\mapsto \sup\limits_{f\in \text{Var}(A)}E[fX]$ has Lebesgue property. Now by relation \ref{2.6} we have:
$$\rho(X_n)=\sup\limits_{a\in A}\langle X_n,a\rangle\leq\sup\limits_{f\in\text{Var}(A)}E[(X_n)^*f]\xrightarrow{n}0.$$
\\Let consider the convex function $\rho$ is not necessarily positive homogeneous. Let $X_n$ and $X$ be bounded in $\mathcal{R}^\infty$ such that $(X_n-X)^*\xrightarrow{\mathbb{P}} 0$. Since $X_n$  is uniformly bounded then there is a bounded sequence of $c_n\in\mathbb{R}^+$ and a positive number $\epsilon$ such that: 
$$\rho(X_n)\leq\sup_{\rho^*(a)\leq c_n}\langle X_n,a\rangle-c_n+\epsilon.$$
Let $c$ be a cluster point of $c_n$ and $I\subseteq\mathbb{N}$ such that $\vert c_n-c\vert<\epsilon$ for all $n\in I$.
\\Let $\rho_1(X):=\sup\limits_{\{\rho^*(a)\leq c+\epsilon\}}\langle X,a\rangle$. Since $\rho_1$ is positively homogeneous , it has Lebesgue property.
Now we have : 

\begin{align*} 
&&\rho(X)\geq\\
&&&\sup_{\{\rho^*(\mu)\leq c+\epsilon\}}\langle X\mu\rangle-c-\epsilon\\
&&&=\rho_1(X)-c-\epsilon\\
&&&=\lim_{n\in I}\rho_1(X_n)-c-\epsilon\\
&&&\geq\lim_{n\in I}\sup_{\rho^*(\mu)\leq c_n}\langle X_n,\mu\rangle-c-\epsilon\\
&&&\geq\lim_{n\in I}\rho(X_n)-3\epsilon\\
&&&\geq\liminf\rho(X_n)-3\epsilon.
\end{align*}
Since $\epsilon>0$ is arbitrary the proof is completed.
\\$\square$

\section{Application in Capital Allocation Problem}
In this section we try to give the application of last section in Capital Allocation Problem with the Fuzzy game approach. This problem for one period coherent risk measures for the first time has been mentioned in earlier work \cite{De2}. In \cite{De2} the author has defined the weak star sub gradient of a coherent risk measure. The author showed that the existence of solution for the capital allocation problem is equivalent to having nonempty sub gradient. By the main theorem of last section we are now able to show that for coherent risk measures , when the capital allocation problem has a solution. 

Here we briefly give the definition of the Capital Allocation Problem , for more details reader is referred to \cite{De2},\cite{Au},\cite{Ar} and \cite{Bi}. Let $X_1,...,X_N$ be $N$ random process in $\mathcal{R}^\infty$ which present $N$ financial items. The total capital required to face the
risk is $\rho(\sum\limits_{i=1}^NX_i)=k$, and we are to find a "fair" allocation $k_1, . . . , k_N$ so that
$k_1 + . . . + k_N = k$. An allocation $k_1, . . . ,k_N$ with $k = k_1 + . . . k_N$  is
called fair if $\forall \alpha_j , j = 1, . . . ,N, 0 \leq \alpha_j \leq 1$ we have: 
$${\sum\limits_j}\alpha_j k_j\leq\rho({\sum\limits_j\alpha X_j}).$$

Now we give the definition and theorems which we need to find a fair allocation.
\begin{definition}
For a function $\rho:\mathcal{R}^\infty\rightarrow \mathbb{R}$ the weak sub-gradient of $\rho$ in $X$ is defined as follows:
\begin{equation}
\bigtriangledown \rho(X):=\{a\in\mathcal{A}^1|\rho(X+Y)\geq\rho(X)+\langle Y,a\rangle\\,\,\,\,\forall Y\in\mathcal{R}^\infty\}.
\end{equation}
This set can be empty. 
\end{definition}
We have the following theorem:
\begin{theorem}
Let $\rho$ be a coherent risk measure on $\mathcal{R}^\infty$ with the Fatou property, given by the
family $\mathcal{P}\subseteq\mathcal{D}_\sigma$. Then $a\in\bigtriangledown\rho(X)$ iff $-a\in\mathcal{P}$ and $\rho(X) = \langle X,g\rangle$.
\end{theorem}
\begin{proof}
Repeat exactly the proof of Theorem 17, Section 8.2 \cite{De2}. 
\\$\square$ 
\end{proof}
{\bf{Remark 5.1}} We should remind that we have the same given definition and theorem for coherent risks on $L^\infty$.

Now we give the following main result of this section which could be interpreted as James's Theorem in our framework:
\begin{theorem}
Let $A$ be a convex, $\sigma(\mathcal{A}^1,\mathcal{R}^\infty)$-closed subset of $\mathcal{D}_\sigma$. $A$ is compact for $\sigma(\mathcal{A}^1,\mathcal{R}^\infty)$ iff for each member $X\in \mathcal{R}^\infty$ it gets its supremum on $A$.  
\end{theorem} 
\begin{proof}
($\Rightarrow $) is obvious.
\\($\Leftarrow $). For the other direction define:
\begin{equation}
\rho(X):=\sup_A\langle X,a\rangle.
\end{equation}
By Theorem 2.1 it is clear that $\rho(-X)$ is a coherent risk with Fatou property. It is not difficult to see that $\text{Var}(A)$ is convex and weak closed subset in $L^1_+$. Let $X\in L^\infty$ be a constant random process. By Remark 3.2 and relation 3.3, $\bar\rho(X)=\sup\limits_{f\in\text{Var}(A)}E[Xf]$. By the assumption of the theorem there exists $a\in A$ such that $\rho(X)=\langle X,a\rangle$ and consequently $\bar\rho(X)=E[\text{Var}(a)X]$. This fact with James's Theorem imply that $\text{Var}(A)$ is weakly compact. Now by Lemma 4.2 we deduce $A$ is compact for topology $\sigma(\mathcal{A}^1,\mathcal{R}^\infty)$.
\\$\square$
\end{proof}
As a direct consequence of Theorems 4.1,5.1 and 5.2 we have :
\begin{theorem}
Let $\rho:\mathcal{R}^\infty\rightarrow \mathbb{R}$ be a coherent risk measure given by $\mathcal{P}\subseteq \mathcal{A}^1$. Then for every $X\in L^\infty$ , $\bigtriangledown\rho(X)\not=\emptyset$ iff $\mathcal{P}$ is $\sigma(\mathcal{A}^1,\mathcal{R}^\infty)$-compact (or $\text{Var}(\mathcal{P})$ is $\sigma(L^1,L^\infty)$-compact) iff $\rho$ (or $\bar{\rho}$) has Lebesgue property.
\end{theorem}

And finally by a simple calculation and using Theorems 5.1,5.2 and 5.3 we have: 
\begin{theorem}
if $X=X_1+...+X_N$ and if $-Q\in\bigtriangledown\rho$ then the allocation $k_i=E_Q[-X_i]$ is a fair allocation .
\end{theorem}

\section{Conclusion.} We have defined the Lebesgue property for convex risk on bounded c\`adl\`ag processes and the static risk due to each convex risk. We have characterized Lebesgue property and we have  shown that $\rho$ has Lebesgue property iff the static risk measure $\bar\rho$ corresponded to that has Lebesgue property. 
This extends the original definition and characterization for the Lebesgue property of \cite{Jo}. Finally we solved the so-called allocation problem in the fair way in the sense of fuzzy game theory. It is our belief that studying risk measures on $\mathcal R^{\infty}$ opens the door to interesting lines of research. 
The coherent risk measures on $\mathcal R^{\infty}$ with the Lebesgue property could be explored in connection with fuzzy game theory problems like those studied in \cite{De}.
\section{Apendix}
Since the proof of theorem 4.2 is a rather technical and long we give the proof in this part.
\\\textbf{Proof of Theorem 4.2.}(2)$\Leftrightarrow $(3). First of all we mention that a subset of $L^p$ is r.c. iff its absolute value is r.c. Actually for $p\neq 1$ this comes from the fact that bounded sets are r.c. sets and for $p=1$, by Dunford-Pettis theorem, uniformly integrable sets are r.c. sets. Let $A_\pm =\{a^\pm |a\in A\}$ where $a^+,a^-$ are increasing decomposition of $a$. It is obvious that $C_\pm =\{(a_T-a_0)^\pm | a\in A\}=\text{Var}(A_\pm )$. So we have $\text{Var}(A)\subseteq2\vert C\vert$ and $C\subseteq2\vert\text{Var}(A)\vert$. Now by above arguments the proof is complete.
\\ 
\\(1)$\Leftrightarrow $(2). We split this part into two cases. 
\\\textbf{Case 1: $p\neq 1$}.
\\Consider $\mathcal{F}_t=\mathcal{F}$. In this case by Theorems 65,67 of Section VII \cite{DM2} we know that $\mathcal{A}^p$ is the dual of $\mathcal{R}^q$. Since $\mathcal{A}^p$ is endowed with the weak* topology then $A$ is r.c. iff it is bounded and this is true iff $\text{Var}(A)$ is bounded or, in the other words, r.c. for topology $\sigma(L^p,L^q)$.
\\When $\mathcal{F}_t$ is nontrivial $A$ is a relatively compact set of $\hat{\mathcal{A}}^p$. The assertion is true because of the continuity of $\Pi^*$.
\\\textbf{Case 2: $p=1$}.
\\($\Rightarrow $):  We claim that $C_\pm $ are relatively compact. Let $a_T^\lambda-a_0^\lambda$ be a net in $C$ and $X$ be a member of $L^\infty$ . Then by the relative compactness of $A$ there is a subnet $a^\beta$ and $a$ such that $a_\beta\xrightarrow{\sigma(\mathcal{A}^1,\mathcal{R}^\infty)} a$. This gives:
 $$E[(a_T^\beta-a_0^\beta)X]=\langle \Pi^{\text{op}}(X),a^\beta\rangle\rightarrow \langle\Pi^{\text{op}}(X),a\rangle=E[(a_T-a_0)X].$$
That means $C$ is r.c. for $\sigma(L^1,L^\infty)$. By Dunford-Pettis theorem we know that this is equivalent to saying that $C$ is uniformly integrable. Then $\vert C\vert=\{\vert f\vert|f\in C\}$ is uniformly integrable and consequently $C_\pm $ are uniformly integrable. Again by Dunford-Pettis $C_\pm $ are r.c.
\\Now by $\text{Var}(A)\subseteq\text{Var}(A_+)+\text{Var}(A_-)=C_++C_-$ we get that $\text{Var}(A)$ is r.c.  
\\($\Leftarrow $):We define a topology on $\mathcal{R}^\infty$. For that we define the semi norms which generate this topology.
\\For any weakly relatively compact subset $H$ in $L^1$ let $V(H):=\{a\in\mathcal{A}^1|\exists f\in H\,\,\text{s.t.}\text{Var}(a)\leq\vert f\vert\}$. Now define the following semi norm for $H$ on $\mathcal{R}^\infty$:
$$P_H(X)=\sup\limits_{a\in V(H)}\langle X,a\rangle.$$
This topology is compatible with the vector structure because obviously the $V(H)$'s are bounded. We show this topology by $\sigma^1$. Let $(\mathcal{R}^\infty)^{'}$ be the dual of $\mathcal{R}^\infty$ with respect to topology $\sigma^1$. It is clear that $\mathcal{A}^1\subseteq (\mathcal{R}^\infty)^{'}$. We want to show that $\mathcal{A}^1= (\mathcal{R}^\infty)^{'}$. 
\\Let $\mu$ be an arbitrarily element of $(\mathcal{R}^\infty)'$ and $X_n$ be a non-negative sequence such that $(X_n)^*\xrightarrow{\mathbb{P}}0 $. Then by Theorem \ref{th:Jo} we have :
\begin{equation}
\label{new}
0\leq P_H(X_n)\leq\sup\limits_{f\in H} E[(X_n)^*\vert f\vert]\rightarrow 0.
\end{equation} 
This gives  $X_n\xrightarrow{\sigma^1} 0$  and then $\mu(X_n)\rightarrow 0$. This fact,and (5.1) of Chapter VII \cite{DM2} show that any $\mu$ can be decomposed into a difference of two positive functionals. Let $\mu^+$ be the positive part. By definition for any $X\geq0$ , $\mu^+(X)=\sup\limits_{0\leq Y\leq X}\mu(Y)$. Let $X_n$ be a positive and decreasing sequence for which $(X_n)^*\downarrow 0$ in probability. Let $0\leq Y_n\leq X_n$ be such that $\mu^+(X_n)\leq \mu(Y_n)+\frac{1}{n}$. Then since $(Y_n)^*\xrightarrow{\mathbb{P}}0 $ by \ref{new} we get: 
$$0\leq\mu^+(X_n)\leq\mu(Y_n)+\frac{1}{n}\rightarrow 0.$$
\\By this fact and Theorem 2 of Chapter VII \cite{DM2} we get that $\mu^+\in\mathcal{A}^1$. Similarly $\mu^-\in\mathcal{A}^1$ so then $\mu\in\mathcal{A}^1$.That means $\mathcal{A}^1=(\mathcal{R}^\infty)'$.
\\The Corollary to Mackey's Theorem 9, Section 13, Chapter 2 \cite{Gr} leads us to $\sigma^1\subseteq\tau(\mathcal{R}^\infty,\mathcal{A}^1)$, where $\tau(\mathcal{R}^\infty,\mathcal{A}^1)$ is the Mackey's topology. By this relation we get that for a relatively weakly compact subset $H$ of $L^1$ there exists $C$, a compact disk in $(\mathcal{A}^1,\sigma(\mathcal{A}^1,\mathcal{R}^\infty))$, for which $\{X| \sup\limits_{a\in C} \langle X,a\rangle< 1\}\subset\{X|P_H(X)\leq1\}$. By polarity $V(H)\subseteq\{X|P_H(X)\leq1\}^\circ\subseteq \{X| \sup\limits_{a\in C} \langle X,a\rangle< 1\}^\circ $. Using the generalized Bourbaki-Alaoglu Theorem we get that $\{X| \sup\limits_{a\in C} \langle X,a\rangle< 1\}^\circ$ is compact in the topology $\sigma(\mathcal{A}^1,\mathcal{R}^\infty)$.
\\Let $H=\text{Var}(A)$. By $A\subseteq V(\text{Var}(A))$,  the proof is complete.  
\\
$\square$

\end{document}